\documentclass[fleqn,twoside]{article}
\usepackage{espcrc2,cite,graphicx,amsmath,amsfonts,amssymb,mathrsfs}
\usepackage[figuresright]{rotating}

\newcommand{\ie}{{\it i.e.}}

\newcommand{\eg}{{\it e.g.}}

\newcommand{\cf}{{\it cf.}}
\newcommand{\etc}{{\it etc.}}
\newcommand{\eq}{Eq.}

\newcommand{\fig}{Figure}

\newcommand{\Ref}{Ref.}
\newcommand{\Refs}{Refs.}

\newcommand{\Tab}{Table}

\newcommand{\stheta}{\sin^22\theta_{13}}
\newcommand{\deltacp}{\delta_{\mathrm{CP}}}
\newcommand{\ldm}{\Delta m_{31}^2}
\newcommand{\sdm}{\Delta m_{21}^2}
\newcommand{\equ}[1]{\eq~(\ref{equ:#1})}
\newcommand{\figu}[1]{\fig~\ref{fig:#1}}

\newcommand{\bi}{\begin{itemize}}
\newcommand{\ei}{\end{itemize}}

\begin{document}

\title{
{\bf Neutrino Oscillation Observables from Mass Matrix Structure}}

\author{{\large Walter Winter},
\address[WUE]{{\it Institut f{\"u}r theoretische Physik und Astrophysik, 
Universit{\"a}t W{\"u}rzburg, D-97074 W{\"urzburg}}}\thanks{E-mail: {\tt winter@physik.uni-wuerzburg.de}}}

\begin{abstract}
\noindent {\bf Abstract}
\vspace{2.5mm}

We present a systematic 
procedure to establish a connection between complex neutrino 
mass matrix textures and experimental observables, including the Dirac CP phase.
In addition,
we illustrate how future experimental measurements affect the selection of textures
in the $(\theta_{13},\deltacp)$-plane. For the mixing angles, we use generic
assumptions motivated by quark-lepton complementarity. 
We allow for any combination between $U_\ell$ and $U_\nu$, as well as
we average over all present complex phases. We find that individual textures
lead to very different distributions of the observables, such as to large or small leptonic
CP violation. In addition, we find that the extended quark-lepton complementarity approach motivates future precision measurements of $\deltacp$ at the level of $\theta_C \simeq 11^\circ$.

\vspace*{0.2cm}
\end{abstract}

\maketitle

{\bf Introduction}. 
By using the same parameterization for $V_{\mathrm{CKM}}$  and $U_{\mathrm{PMNS}}$, and by quantifying
the differences between these two mixing matrices, it is implied that the quark and lepton
sectors might be somehow connected. Recently, interesting ``quark-lepton
complementarity'' (QLC) relations~\cite{Petcov:1993rk,Smirnov:2004ju,Raidal:2004iw,Minakata:2004xt} 
have been proposed, which could be indicative
for such a quark-lepton unification. These QLC relations suggest
empirical connections between quark and lepton mixings, such as
\begin{equation}\label{equ:qlc}
\theta_{12}+\theta_\text{C} \simeq \pi/4  \simeq \theta_{23} \, . 
\end{equation}
A simple underlying hypothesis may therefore be
that all mixings in the charged lepton and neutrino sectors 
are either maximal, or Cabibbo-like. It can be motivated by the observation 
that mixing angles $\sim \theta_\text{C}$ (and powers $\sim\theta_\text{C}^n$ thereof), 
as well as maximal mixing, can be readily obtained in models from flavor symmetries.
Consequently, any deviation from maximal mixing in the 
large leptonic mixing angles $\theta_{12}$
and $\theta_{23}$ can only arise as a result of taking the product of the charged
lepton mixing matrix $U_\ell$ and the neutrino mixing matrix
$U_\nu$ in the PMNS mixing matrix
\begin{equation}
U_{\rm PMNS}=U_\ell^\dagger U_\nu \, .
\label{equ:upmns}
\end{equation}
If one assumes that all mixing angles in $U_\ell$ and $U_\nu$ can only be from the
sequence $\{\pi/4, 0, \epsilon, \epsilon^2, \hdots \}$ with $\epsilon \simeq \theta_C$,
one can systematically construct the parameter space of all possible combinations of
$U_\ell$ and $U_\nu$ in this framework and choose the realizations being compatible 
with data. This analysis was performed in \Ref~\cite{Plentinger:2006nb} for the
case of real matrices up to oder $\epsilon^2$, and it was called ``extended quark-lepton
complementarity'' (for a seesaw implementation, see \Ref~\cite{Plentinger:2007px}). Simple conventional quark-lepton complementarity
implementations, such as $U_{\rm PMNS} \simeq V_{\rm CKM}^\dagger U_{\rm bimax}$, emerge as special cases
in this approach (see, \eg, \Refs~\cite{Datta:2005ci,Everett:2005ku}), but are not the exclusive solutions. For example, the charged lepton
sector may actually induce two large mixing angles. 

Any realization of \equ{upmns} can be used to construct the effective 
Majorana neutrino mass matrix as
\begin{equation}
M_\nu^{\rm Maj} = U_\nu M_\nu^{\rm diag} U_\nu^T \, ,
\label{equ:maj}
\end{equation}
where we can use the experimentally motivated mass eigenvalues, such as $m_1:m_2:m_3 \simeq \epsilon^2:\epsilon:1$
for the normal hierarchy. This additional assumption for the neutrino mass eigenvalues is
compatible with the current measurements of $\sdm$ and $\ldm$.
Similarly, the charged lepton
and quark mass hierarchies can be described by powers of $\epsilon$ as well, which means that
by our hypothesis, all mixings and hierarchies are induced by a single small quantity $\epsilon \simeq \theta_C$ as a potential remnant of a unified theory. By identifying the leading order entries in the mass matrix realization \equ{maj}, the ``texture'', one can establish a connection to theoretical models.
For example, masses for quarks and leptons may arise from higher-dimension terms via
the Froggatt-Nielsen mechanism~\cite{Froggatt:1978nt} in combination with a flavor symmetry:
\begin{equation}
\mathcal{L}_\mathrm{eff} \sim \langle H \rangle \, \epsilon^n \, \bar{\Psi}_L \Psi_R \, .
\label{equ:fn}
\end{equation}
In this case, $\epsilon$ becomes meaningful in terms of a small
parameter $\epsilon=v/M_F$ which controls the flavor symmetry breaking.\footnote{
Here $v$ are universal VEVs of SM singlet scalar ``flavons'' that break the flavor symmetry,
and $M_F$ refers to the mass of superheavy fermions, which are charged under the
flavor symmetry. The SM fermions are given by the $\Psi$'s.} 
The integer power of $\epsilon$ is solely determined by the
quantum numbers of the fermions under the flavor symmetry (see, \eg, \Refs~\cite{Enkhbat:2005xb,Plentinger:2007px}). 

In this letter, we demonstrate how one can construct the full {\em complex} parameter space of realizations of \equ{upmns} from generic assumptions. We use the context of extended quark-lepton complementarity
to illustrate our procedure, where we average over all possible complex phases. Since we will obtain a $1:n$ correspondence between a texture and a number of valid (experimentally allowed) realizations of this texture, we can study the distributions of observables corresponding to the realizations.  We will focus
on the effective Majorana neutrino mass matrix for the normal hierarchy, but this procedure can easily
be extended to the neutrino Dirac mass matrix and charged lepton mass matrix, as well as one can use different neutrino mass schemes~\cite{Plentinger:2006nb}, or different generic input assumptions.

{\bf Method. }Following the procedure in \Ref~\cite{Plentinger:2006nb}, the PMNS matrix can, in general,
be  written as the product of two matrices in the CKM-like standard parameterization $\widehat{U}$:
\begin{equation}\label{equ:pmnspara}
U_\text{PMNS}=\widehat{U}_\ell^\dagger U_\nu=\widehat{U}_\ell^\dagger D\widehat{U}_\nu\,K,
\end{equation}
Here
$D=\text{diag}(1,e^{\text{i} \varphi_1},e^{\text{i} \varphi_2})$
and $K=\text{diag}(e^{\text{i} \phi_1},e^{\text{i} \phi_2},1)$ are
remaining diagonal matrices with phases in the range
$\varphi_1,\varphi_2,\phi_1,\phi_2\in[0,2\pi)$,
which cannot be rotated away in general because of the CKM-like parameterizations of $\widehat{U}_\ell$ and
$\widehat{U}_\nu$. In addition,  $\widehat{U}_\alpha$ can be parameterized by
three mixing angles $\theta_{12}^\alpha$, $\theta_{13}^\alpha$, and $\theta_{23}^\alpha$,
as well as one Dirac-like phase $\delta^\alpha$ in the usual way.
Our three-step procedure then reads:

{\bf Step~1}
We generate all possible pairs $\{ U_\ell$, $U_\nu \}$ described by
\begin{equation}
\{\theta_{12}^\ell,\theta_{13}^\ell,\theta_{23}^\ell,\delta^\ell,\theta_{12}^\nu,\theta_{13}^\nu,\theta_{23}^\nu,\delta^\nu,\varphi_1,\varphi_2,\phi_1,\phi_2\} \nonumber
\end{equation}
with $\sin \theta_{ij}^\alpha \in \{ 1/\sqrt{2}, \epsilon, \epsilon^2, 0 \}$ (cut off by the
current experimental precision) and 
uniform distributions of all phases in 32 steps each, which corresponds to an
averaging over the phases.\footnote{We have checked that 32 steps are sufficient to
reproduce the general qualitative features.} 

{\bf Step~2}
Then we calculate $U_{\rm PMNS}$ by \equ{upmns}, read off the mixing angles and
physical phases, and select those
{\em realizations} with mixing angles being compatible with current data at 
the $3 \sigma$ confidence level (\cf, \Ref~\cite{Plentinger:2006nb}).

{\bf Step~3}
For each valid realization, we find the corresponding {\em texture} by 
computing \equ{maj}, expanding in $\epsilon$, and by
identifying the first non-vanishing coefficient, which is leading to the
texture entry $1$, $\epsilon$, $\epsilon^2$, or $0$.\footnote{Note that this definition of a texture
only includes the absolute value of the leading coefficient, while more sophisticated
concepts may include the phase as well.} 

Obviously, many possible realizations
will lead to the same texture, \ie, there will be a $1:n$ correspondence between
textures and realizations. We will therefore show the distributions of observables
for all valid realizations (valid choices of order one coefficients) leading to a specific texture. 
The interpretation of the results has then to be done in the reverse direction:
A certain model will lead to a specific texture,
which can be fit to data by choosing the order one coefficients from our set of realizations. 
Note that the realizations connect to experimental observations, while the textures connect to theoretical models.

\begin{table*}[p]
\begin{center}
\begin{tabular}{cccl}
\hline
No. & Texture & Percentage & Distributions of valid realizations leading to this texture \\
\hline
\#1 &  $\left(
\begin{array}{lll}
 \epsilon  & \epsilon  & \epsilon  \\
 \epsilon  & 1 & 1 \\
 \epsilon  & 1 & 1
\end{array}
\right)$ & 41\% & \raisebox{-1cm}{\includegraphics[width=9cm]{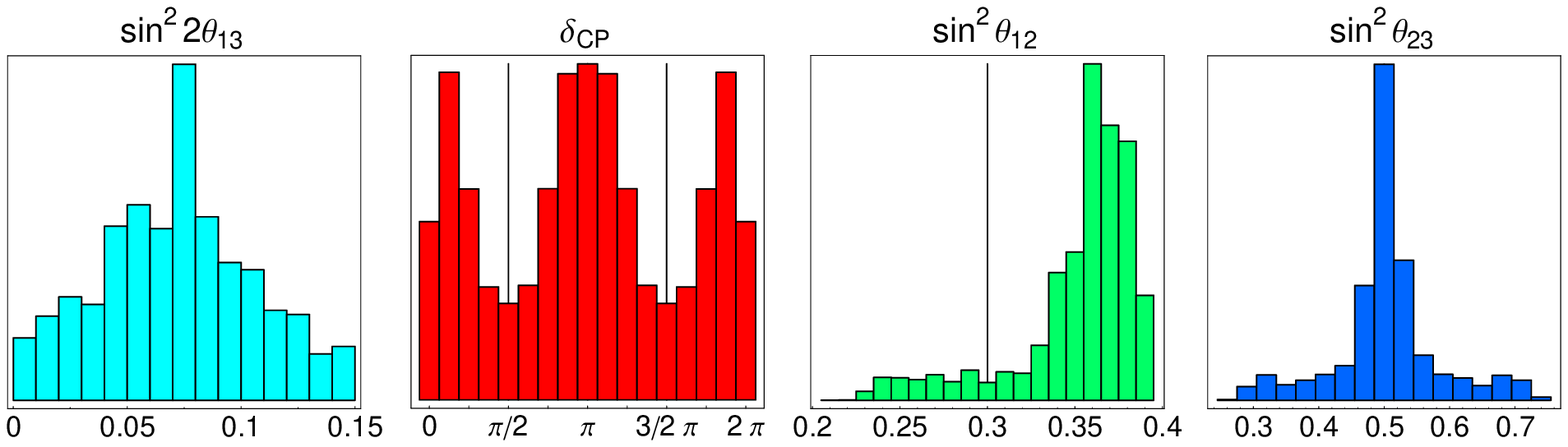}} \\ 
\hline
\#2 & $\left(
\begin{array}{lll}
 \epsilon  & \epsilon  & \epsilon ^2 \\
 \epsilon  & \epsilon  & \epsilon  \\
 \epsilon ^2 & \epsilon  & 1
\end{array}
\right)$
 & 8.4\% & \raisebox{-1cm}{\includegraphics[width=9cm]{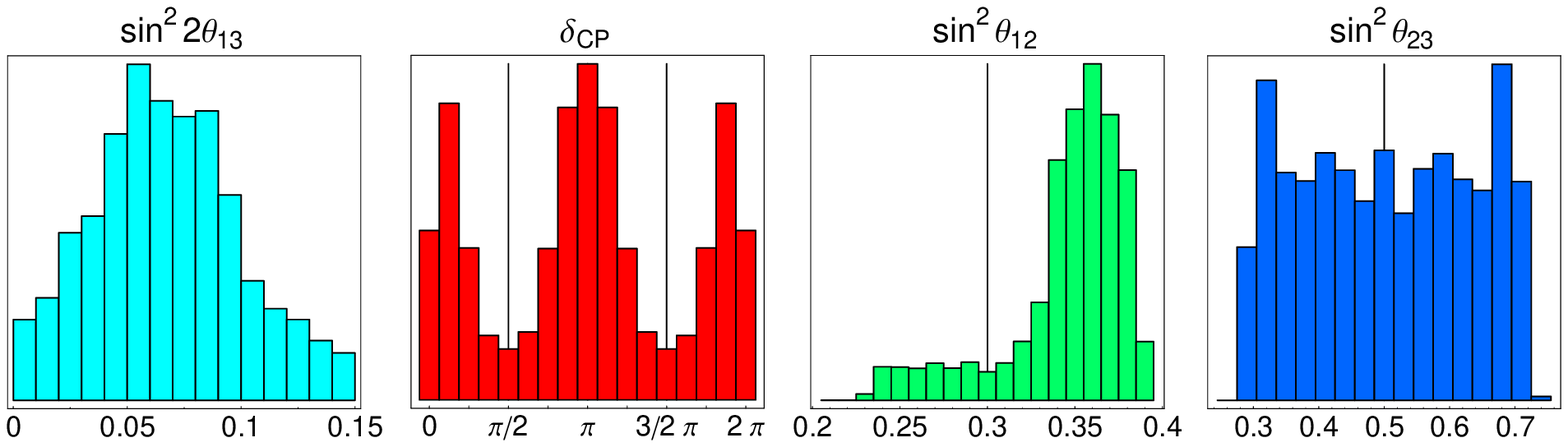}} \\ 
\hline
\#3 &$\left(
\begin{array}{lll}
 \epsilon  & \epsilon  & 0 \\
 \epsilon  & \epsilon  & 0 \\
 0 & 0 & 1
\end{array}
\right)$ & 5.6\% & \raisebox{-1cm}{\includegraphics[width=9cm]{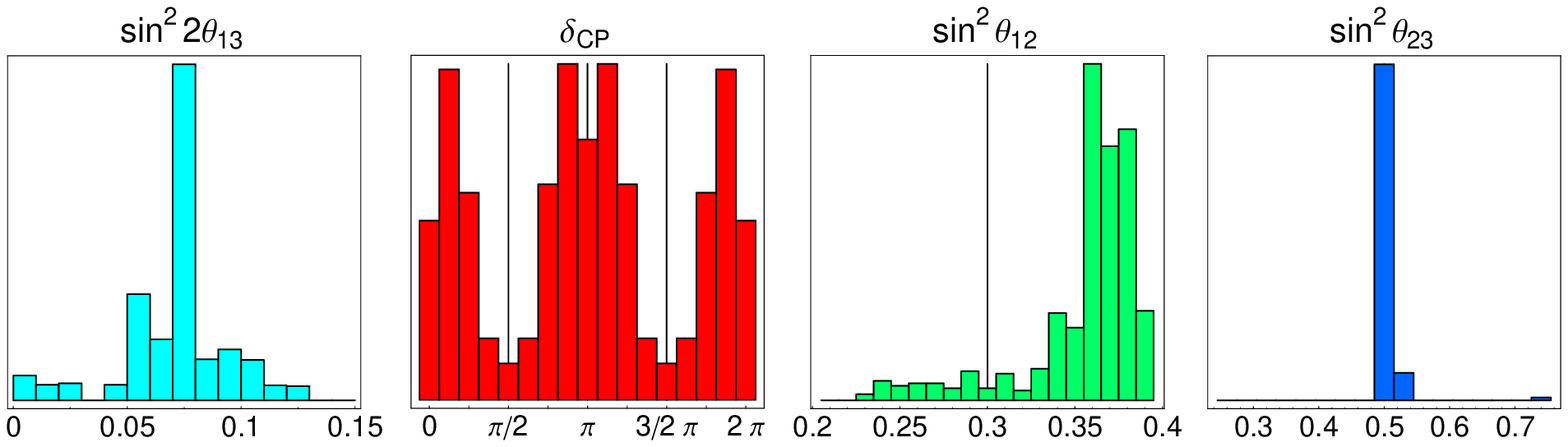}} \\ 
\hline
\#4 & $\left(
\begin{array}{lll}
 1 & \epsilon ^2 & 1 \\
 \epsilon ^2 & \epsilon  & \epsilon ^2 \\
 1 & \epsilon ^2 & 1
\end{array}
\right)$ & 5.4\% & \raisebox{-1cm}{\includegraphics[width=9cm]{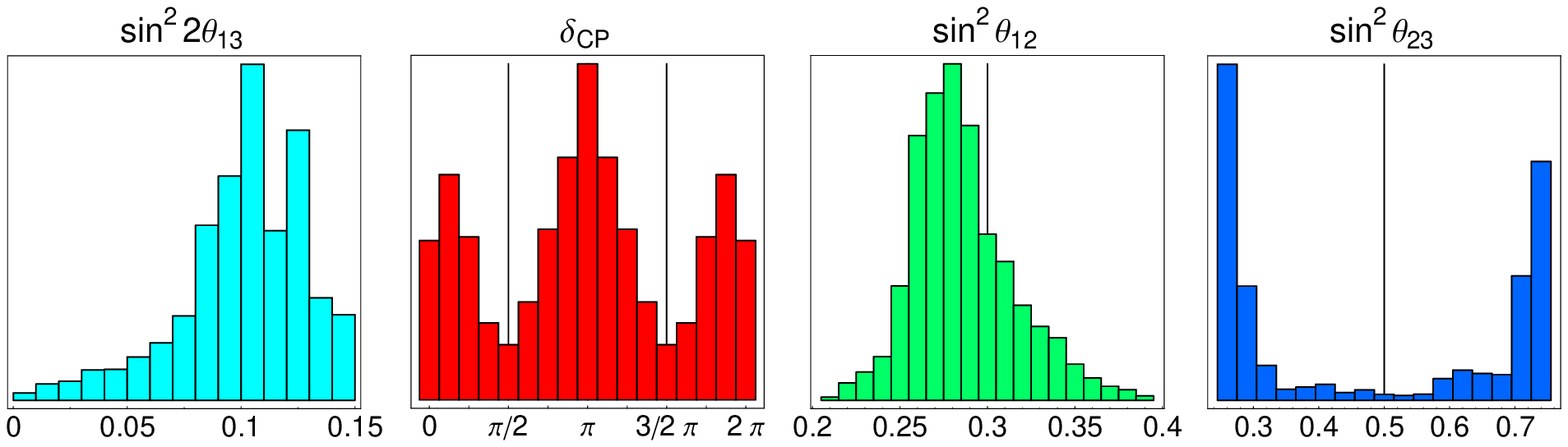}} \\ 
\hline
\#5  & $\left(
\begin{array}{lll}
 \epsilon  & \epsilon  & \epsilon  \\
 \epsilon  & \epsilon  & \epsilon ^2 \\
 \epsilon  & \epsilon ^2 & 1
\end{array}
\right)$ & 4.8\% & \raisebox{-1cm}{\includegraphics[width=9cm]{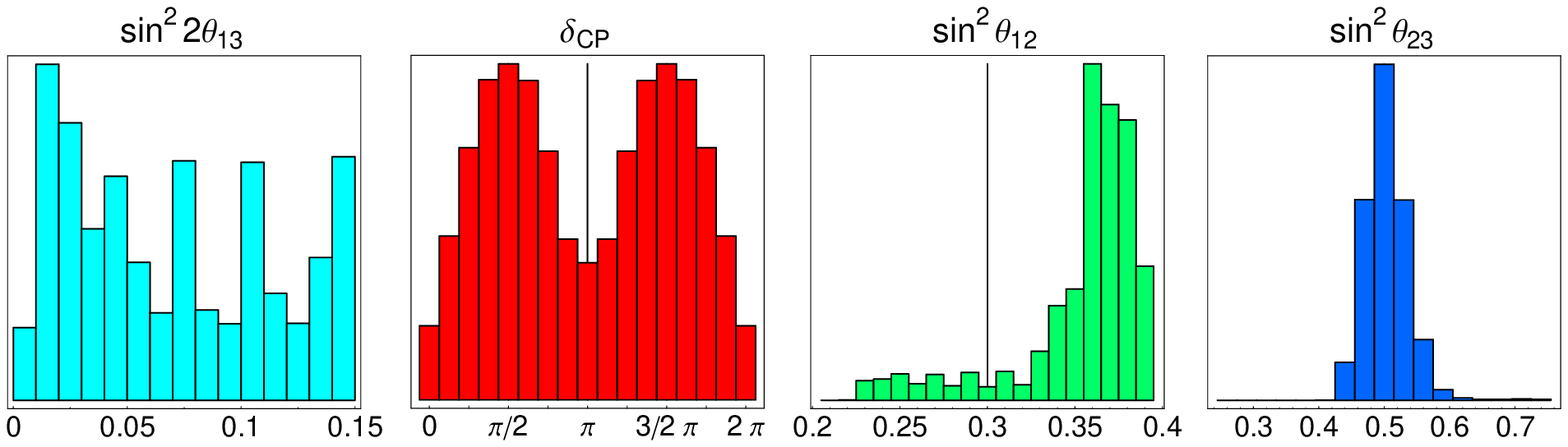}} \\ 
\hline
\#6 & $\left(
\begin{array}{lll}
 1 & 0 & 1 \\
 0 & \epsilon  & 0 \\
 1 & 0 & 1
\end{array}
\right)$ & 1.4\% & \raisebox{-1cm}{\includegraphics[width=9cm]{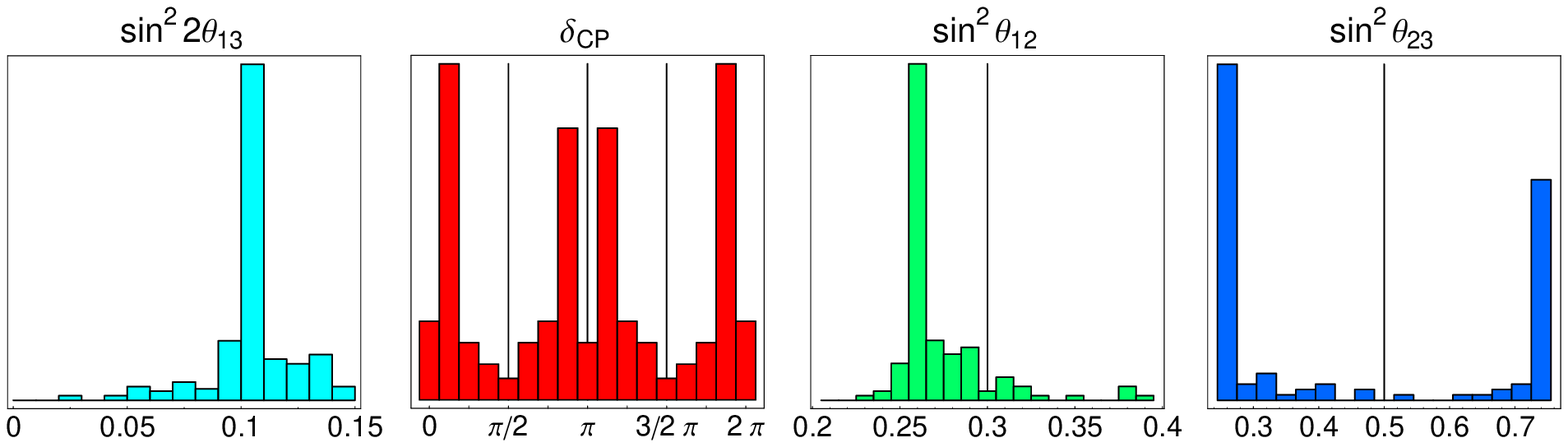}} \\ 
\hline
\#7 & $\left(
\begin{array}{lll}
 \epsilon ^2 & \epsilon ^2 & \epsilon ^2 \\
 \epsilon ^2 & 1 & 1 \\
 \epsilon ^2 & 1 & 1
\end{array}
\right)$ & 0.5\% & \raisebox{-1cm}{\includegraphics[width=9cm]{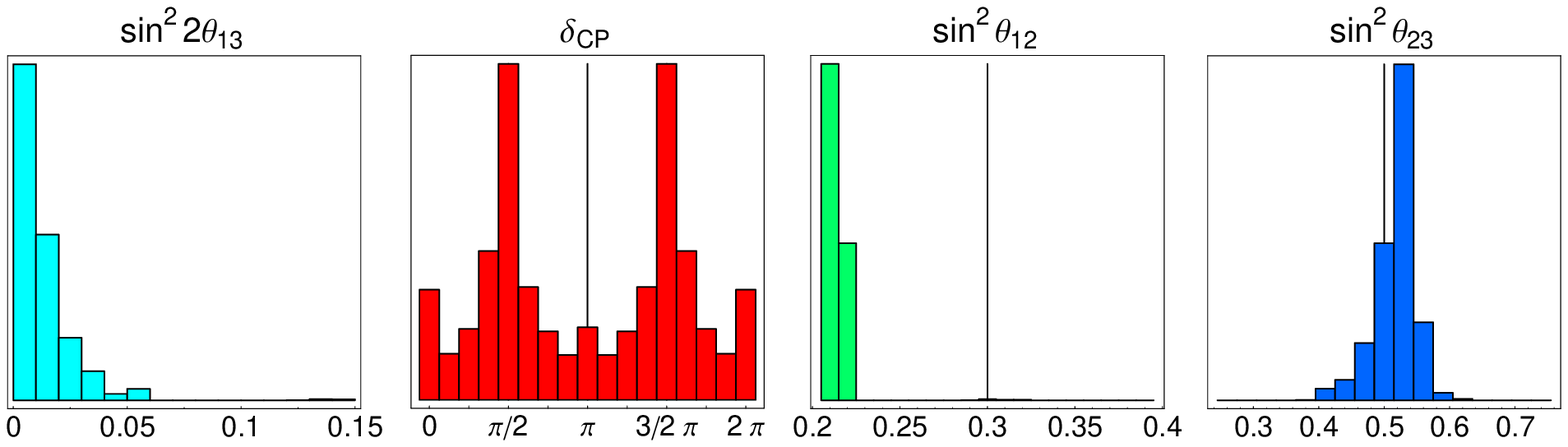}} \\ 
\hline
\end{tabular}
\end{center}
\caption{\label{tab:textures}Selected textures for $M_\nu^{\rm Maj}$. ``Percentage'' refers to the
fraction of all realizations leading to a texture.}
\end{table*}

\begin{table*}[t]
\begin{center}
\begin{tabular}{cllll}
\hline
Text. & Observables & \multicolumn{3}{c}{Input parameters} \\
No. &   $(\sin^2 \theta_{12},\sin^2 2 \theta_{13}, \sin^2 \theta_{23}, \delta)$ &
$(s_{12}^\ell,s_{13}^\ell, s_{23}^\ell, \delta^\ell)$ &
$(s_{12}^\nu,s_{13}^\nu, s_{23}^\nu, \delta^\nu)$ & $(\varphi_1,\varphi_2,\phi_1,\phi_2)$ \\  
\hline
\#1 & $(0.30,0.15,0.50,4.70)$ & $(\epsilon , \epsilon , \epsilon ^2, 5.30)$ & 
$(\frac{1}{\sqrt{2}}, \epsilon , \frac{1}{\sqrt{2}}, 4.71)$ &  $(5.50, 1.37, 0, 0)$  \\
\#2  & $(0.30, 0.15, 0.50,4.74)$ & $(\epsilon , \epsilon , \frac{1}{\sqrt{2}}, 4.32)$ & 
$(\frac{1}{\sqrt{2}},0, \epsilon , 0)$ & $(5.30, 0.59, 0, 0)$ \\ 
\#3  & $(0.30, 0.15, 0.50,4.74)$ & $(\epsilon , \epsilon , \frac{1}{\sqrt{2}}, 4.71)$ & 
$(\frac{1}{\sqrt{2}},0, 0 , 0)$ & $(5.50, 2.55, 0, 0)$ \\ 
\#4  & $(0.30, 0.01, 0.50,2.61)$ & $(\frac{1}{\sqrt{2}},\frac{1}{\sqrt{2}},\frac{1}{\sqrt{2}},4.91)$ & 
$(\epsilon,\frac{1}{\sqrt{2}},\epsilon^2,0.20)$ & $(0.20, 3.53, 0, 0)$ \\ 
\#5 & $(0.30, 0.05, 0.50,4.89)$ & $(\epsilon,\epsilon,\frac{1}{\sqrt{2}},0.59)$ & 
$(\frac{1}{\sqrt{2}},\epsilon,\epsilon^2,0.39)$ & $(5.89, 1.18, 0, 0)$ \\ 
\#6 & $(0.32, 0.13, 0.47,3.57)$ & $(\frac{1}{\sqrt{2}},\frac{1}{\sqrt{2}},\frac{1}{\sqrt{2}},1.77)$ & 
$(0,\frac{1}{\sqrt{2}},0,3.53)$ & $(0,5.69, 0, 0)$ \\ 
\#7  & $(0.22, 0.01, 0.50,3.18)$ & $(\epsilon,\epsilon,\epsilon^2,2.55)$ & 
$(\epsilon,\epsilon^2,\frac{1}{\sqrt{2}},3.14)$ & $(3.14,2.55, 0, 0)$ \\ 
\hline
\end{tabular}
\end{center}
\caption{\label{tab:real}Examples for specific realizations for the textures in \Tab~\ref{tab:textures}
(including the mixings from the lepton sector). All of the shown realizations have observables very close to the current best-fit values.}
\end{table*}

{\bf Results and interpretation. }
We find 29 different textures for $M_\nu^{\rm Maj}$ from the above procedure. These
come from realizations which represent valid choices of
$U_{\mathrm{PMNS}}$. In \Tab~\ref{tab:textures}, we show seven selected textures (see \Ref~\cite{CTexWebpage}
for a complete list and examples of corresponding $M_\ell$ textures computed as in \Ref~\cite{Plentinger:2006nb}). The percentage
of all realizations leading to a specific texture is given in the third column.
The distributions of the observables $\sin^2 2 \theta_{13}$, $\delta$, $\sin^2 \theta_{12}$
and $\sin^2 \theta_{23}$ are given in the last column (arbitrary units), where the current best-fit
values are marked by vertical lines. In \Tab~\ref{tab:real}, we give specific examples
for realizations leading to a specific texture, where we have chosen cases with
$\sin^2 \theta_{12}$ and $\sin^2 \theta_{23}$ very close to the current best-fit values.
One can read off this table valid combinations between $U_\ell$ and $U_\nu$ leading to specific
textures referred to by the texture number. For example, only the shown realizations leading to textures \#1 and \#7 have exclusively small mixings coming from the lepton sector.
The choice $\phi_1=\phi_2=0$ in this table is 
accidental (it does not appear for more seldom textures).
In comparison to \Ref~\cite{Plentinger:2006nb}, we focus on $M_\nu^{\rm Maj}$, and 
we obtain a much larger number of textures for $M_\nu^{\rm Maj}$. Allowing for the current
measurement errors instead of the more stringent ones used in \Ref~\cite{Plentinger:2006nb},
13 more textures in addition to the 6 original ones are allowed. If, in addition, complex phases are 
introduced instead of using the real case only, we find 10 more textures, leading to a total of 29 (\cf, \Ref~\cite{CTexWebpage} for details on which texturs are falling into which category). 

As far as the interpretation of the distributions of the observables in \Tab~\ref{tab:textures} is concerned,
it certainly depends on the measure of the input parameter space, in particular, our choice of discrete values for the mixing angles. Our choice $\propto \epsilon^n$ corresponds to a uniform (anarchic) distribution on a logarithmic scale. At the mass matrix level, one may justify such an assumption
by the Froggatt-Nielson mechanism by using an arbitrary number of heavy fermion propagators 
in \equ{fn}. The quark and lepton mass hierarchies seem to obey such a ``logarithmic uniformity'' as well, which means
that this assumption may be well motivated for the eigenvalues. Our choices for the mixing angles correspond to the postulate that
we find, at least roughly, this distribution in the matrix elements reflected in the mixing angles. We have therefore checked that our distribution of mixing angles translates into a similar (uniform) distribution
of mass matrix powers if one allows for all possible hierarchies. In fact, there is a slight (but not order of magnitude-wise) deviation from this uniformity for the diagonal elements leading to a peak at $1$, and  for the off-diagonal elements leading to a peak at $\epsilon^2$. The choice of a normal
neutrino mass hierarchy is an additional, experimentally motivated constraint, which obviously affects the mapping between the mass matrix and mixing angle parameter spaces.

Given our assumptions for the input values, the interpretation of our figures is then as follows from
the experimental point of view:
For the valleys where no realizations are found, a measurement could exclude a texture. For 
the peaks, where most realizations are found, a measurement would confirm the most ``natural'' choice of observables. This naturalness argument may be similar to a landscape interpretation, such as in \Refs~\cite{Hall:2007zj,Hall:2007zh} (using a different measure). Note, however, that we impose at least
some flavor structure before we obtain the distributions. 
Similarly, the figures can be interpreted from the model building point of view: The peaks correspond to plenty of possibilities how a specific texture can be implemented. For the valleys, where only a few realizations are found, exceptional realizations can lead to this texture. This means that one can basically read off such tables which textures to use if one wants to produce small $\stheta$, large $\stheta$, deviations from maximal mixing, \etc.

We discuss now the different observables. For $\sin^2 \theta_{23}$, either maximal mixing, or relatively
large deviations from maximal mixing can be observed. Only texture \#2 has a relatively broad distribution in this observable. Deviations from maximal mixing will therefore be
an important model discriminator, see \Ref~\cite{Antusch:2004yx}.  For $\sin^2 \theta_{12}$, the current best-fit value can only be exactly generated in very few cases, which is not surprising since $\theta_{12}$ has to emerge from combinations between maximal mixing and $\theta_C$ in our approach. 
Only texture \#4 covers the current best-fit value very well. Therefore, precise measurements of $\sin^2 \theta_{12}$ will be very valuable for such a quark-lepton complementarity ansatz.
For $\stheta$, any case can be found: large $\stheta$ (\eg, \#4), small $\stheta$ (\eg, \#7), medium $\stheta$ (\eg, \#1), or a broad distribution in $\stheta$ (\eg, \#5). 
And for $\deltacp$, maximal CP violation (\#5 and \#7), CP conservation, or small deviations from these cases at the level of $\pi/16 \sim \theta_C$ are present.
 This can be understood as follows: The phases may be given in the symmetry base of an underlying theory, where uniform distributions (or any other assumptions) may be well motivated. These assumptions translate (via invariants) into the observables, where combinations with the mixing angles enter. Obviously, by choosing powers of the Cabibbo angle for the mixing angles, these powers will somehow translate into the phase distributions.
This implies that a measurement precision of $\sim 11^\circ$ 
may be a reasonable requirement for future experiments to test a small CP violation. Such a precision could be obtained in optimized beta beams or neutrino factories (see, \eg, \Ref~\cite{Huber:2004gg}). 

%\section{Impact of Future ($\boldsymbol{\theta_{13},\delta}$) Measurements}

\begin{figure*}[t!]
\begin{center}
\includegraphics[width=0.8\textwidth]{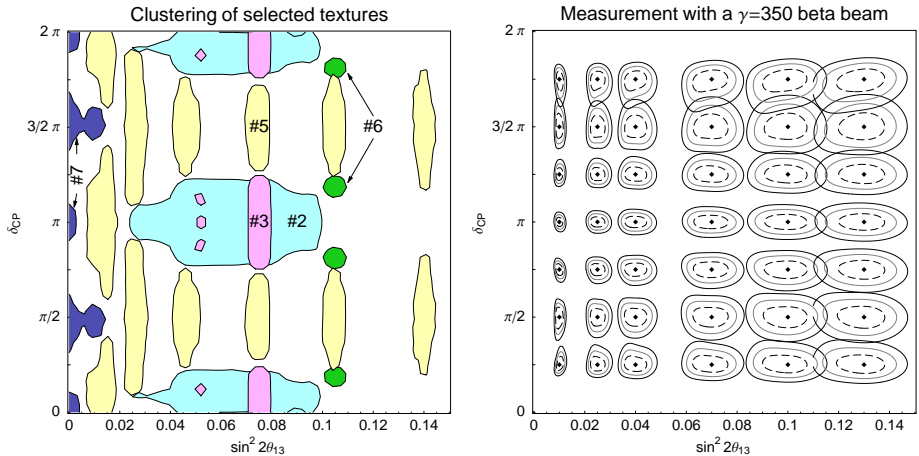}

\vspace*{-1cm}

\end{center}
\caption{\label{fig:th13delta} Left panel: Clustering of specific textures (as labeled in the plot) in the $\stheta$-$\deltacp$-plane. The clusters contain 50\% of all realizations leading to a specific texture.
Right panel: Measurement precision of a $\gamma=350$ beta beam for different selected true values
of $\stheta$ and $\deltacp$ (diamonds). The contours correspond to $1\sigma$, $2\sigma$, and $3\sigma$ (2 d.o.f., not shown oscillation parameters marginalized over).}
\end{figure*}

Let us now discuss the use of simultaneous constraints on $\stheta$ and $\deltacp$ on this texture
space. We show in \figu{th13delta}, left, clusters representing
50\% of all valid realizations for different selected textures in the $\stheta$-$\deltacp$-plane.
In \figu{th13delta}, right, we show the precision for a typical potential future high
precision instrument, namely the 
$\gamma=350$ beta beam option from \Ref~\cite{Burguet-Castell:2005pa}
simulated with GLoBES~\cite{Huber:2004ka,Huber:2007ji}.\footnote{This setup assumes 
eight years of simultaneous
  operation with $2.9 \cdot 10^{18}$ useful $^6$He and $1.1 \, \cdot
  10^{18}$ useful $^{18}$Ne decays per year and a $500 \, \mathrm{kt}$
  water Cherenkov detector. The gamma factor is $350$ for both
  isotopes, and the baseline is $L=730 \, \mathrm{km}$. The setups is
  simulated with the migration matrixes from
  \Ref~\cite{Burguet-Castell:2005pa}. In order to impose constraints
  to the atmospheric parameters, ten years of T2K disappearance
  information is added (such as in \Ref~\cite{Huber:2005jk}).}
From \figu{th13delta}, left, which represents the two-dimensional version of the histograms in
\Tab~\ref{tab:textures}, we find that the realization distributions for 
different textures cluster in different, often non-overlapping regions of the $\stheta$-$\deltacp$-plane.  
Therefore, in order to distinguish many different textures, combined
information on both $\stheta$ and $\deltacp$ is useful. For example,
for \#5, all possible values of $\stheta$ and $\deltacp$ are covered by the clusters. However, a
simultaneous measurement of $\stheta$ and $\deltacp$ (\cf, right panel) may easily indicate
that this cluster is not realized because certain regions in the $\stheta$-$\deltacp$-plane are sparsely
populated.  From the right panel we learn that, compared to the texture cluster sizes, the future experiments will provide very precise measurements in this parameter space. In particular cases, such as for texture \#6, a separate measurement of $\stheta$ or $\deltacp$ could hardly exclude the texture, but a 
combined measurement might (depending on the best-fit point). Therefore, we conclude that a simultaneous measurement of $\stheta$ and $\deltacp$, as it is usually discussed from the experimental point of view, will be a much stronger discriminator than an individual measurement of one of these parameters.

{\bf Summary and conclusions.} We have demonstrated how one can systematically construct experimentally allowed realizations of $U_{\mathrm{PMNS}}$ from very generic assumptions, and we have used them to relate neutrino mass textures with observables. Our procedure can be easily applied to other observables, different (or additional) assumptions, different neutrino mass schemes, or to the Dirac mass matrix case. The resulting distributions of observables could be useful for experiments, such as to illustrate their exclusion power in the model space, and theorists, such as to identify textures connected with specific distributions for the observables. It turns out the possibilities leading to a specific texture can often be connected with very characteristic observable distributions, such as small $\stheta$, large $\stheta$, strong CP violation, CP conservation, a strong deviation from maximal mixing, \etc.

As an example, we have used the context of extended quark-lepton complementarity to apply our procedure:
All mixing angles are forced to either zero or maximal by a symmetry, or are generated by a quantity $\simeq \theta_C$ as a single remnant from a Grand Unified Theory. This framework means that the solar mixing angle can only emerge as a combination between maximal mixing and the Cabibbo angle, and it is directly related to the quark sector. As more specific conclusions from this assumption, we find that $\theta_{12}$ will be an important indicator for specific textures, \ie, future precision measurements of $\theta_{12}$ will be very selective. In addition, the extended quark-lepton complementarity ansatz motivates future precision
measurements of $\deltacp$ at the level of $\pm \theta_C \simeq \pm \pi/16 \simeq 11^\circ$. Such hints are important for the design of future experiments, since one would like to know how far one has to go experimentally.

{\bf Acknowledgments:}
I would like to thank J{\"o}rn Kersten, Hitoshi Murayama, Florian Plentinger, and Gerhart Seidl
for useful discussions. This work has been supported by the Emmy Noether program of
DFG.

% \vspace{-0.5cm}

% \bibliography{references}
% \bibliographystyle{h-elsevier}

\end{document}